\documentclass{article}[10pt]
\usepackage{epsfig}
\def\neane{\buildrel {(-)} \over {\nu}}

\newcommand{\cm}{\rm \,cm}

\newcommand{\m}{\rm \,m}

\newcommand{\km}{\rm \,km}

\newcommand{\s}{\rm \,s}

\newcommand{\sr}{\rm \,sr}

\newcommand{\MeV}{\rm \,MeV}
\newcommand{\eV}{\rm \,eV}

\newcommand{\GeV}{\rm \,GeV}
\newcommand{\TeV}{\rm \,TeV}

\newcommand{\lsim}{\lower .5ex\hbox{$\buildrel < \over {\sim}$}}
\newcommand{\gsim}{\lower .5ex\hbox{$\buildrel > \over {\sim}$}}

\begin{document}
{\bf\Large \hskip 11cm DFUB 00/17 \\}
\vskip 2. cm
\begin{center}
{\bf\Large Neutrino physics and astrophysics with the MACRO detector}
\vskip 0.4 cm
{\bf \large Invited paper at the Chacaltaya Meeting on Cosmic Ray Physics, \\
La Paz,
23-27 July, 2000\\}
\vskip 0.4 cm
{\large\noindent G. Giacomelli and A. Margiotta \\
\vskip 0.2 cm
Dipartimento di Fisica dell'Universit\`a di Bologna 
and INFN, I-40127 Bologna, Italy}
\end{center}
\begin{abstract}
After a brief presentation of the MACRO detector we discuss the data on
atmospheric neutrinos and neutrino oscillations, on high energy 
($E_\nu > 1 \ \GeV$) neutrino astronomy, on indirect searches for WIMPs and 
low energy ($E_\nu \gsim 7 \ \MeV$) stellar collapse neutrinos.
\end{abstract}

\section{Introduction}
MACRO is a multipurpose underground detector designed to search
for rare events in the cosmic radiation. It was optimized to search for
the supermassive magnetic monopoles (MMs) predicted by Grand Unified Theories (GUT)
of the electroweak and strong interactions.
MACRO performs measurements in areas of astrophysics, nuclear, particle and 
cosmic ray physics. In this paper we shall discuss atmospheric 
neutrinos and neutrino oscillations, high energy $(E_\nu \ \gsim\ 1\ \GeV)$ 
neutrino astronomy, indirect searches for WIMPs and searches for low energy ($E_\nu \ 
\gsim\ 7\  $MeV) stellar collapse neutrinos.
The  detector has global dimensions of $12\times9.3\times76.5~{\m}^3$ and 
provides a total acceptance to an isotropic flux of particles of $\sim 
10,000 {\m}^2{\sr}$. The total mass is $\simeq 5300 \ $t.
A cross section of the  detector is shown in Fig. \ref{cross_det} 
\cite{ncim86}. 
It has three sub-detectors: liquid scintillation counters, 
limited streamer tubes and nuclear track detectors. The mean rock depth of the 
overburden is $\simeq 3700~ $m.w.e.; the 
minimum is $3150~ $ m.w.e. This defines the minimum downgoing muon energy at 
the surface 
as $\sim 1.3 {\TeV}$ in order to reach the underground lab.
The average residual energy and the muon flux at the lab depth are
$\sim 310~{\GeV}$ and $\sim 1~{\m}^{-2}~{\rm h}^{-1}$, respectively. 

\section{Atmospheric $\nu$. Neutrino oscillations.}
{\bf{Neutrino oscillations.}}
For massive neutrinos one has to consider the weak flavour eigenstates 
($\nu_e$, $\nu_\mu$, $\nu_\tau$, which are relevant for $\pi \rightarrow 
\mu \nu_\mu$, K$\rightarrow \mu \nu_\mu$ and
 $\mu \rightarrow e \nu_e \nu_\mu$ decays and for $\nu$ interactions) and mass eigenstates ($\nu_1$, $\nu_2$, 
$\nu_3$, relevant for propagation). The flavour eigenstates are linear 
combinations of the mass 
eigenstates, \( \nu_l = \sum_{m=1}^{3} U_{lm} \nu_m\).
For two flavours
\begin{equation}
  \left\{ \begin{array}{l}\nu_\mu = \nu_2 cos\theta + \nu_3 sin\theta \\
  \nu_\tau = -\nu_2 sin\theta + \nu_3 cos\theta 
\end{array}
\right. 
\end{equation}
where $\theta $ is the mixing angle. The rate of $\nu_\mu$ disappearance is
given by 
\begin{equation} P(\nu_\mu \rightarrow \nu_\mu) \simeq 1 - sin^2 2\theta \, sin^2 
(1.27 \Delta m^2  L / E_\nu) \end{equation}
$\Delta m^2 = {m^2}_{\nu_3} - {m^2}_{\nu_2} $, {\it L} = distance (in m) from the 
decay point to the interaction point.

Only disappearance experiments have been performed until now; the final proof 
of $\nu$-oscillations will come from appearance experiments, 
$ P(\nu_\mu \rightarrow \nu_\tau) = 1 - P(\nu_\mu \rightarrow \nu_\mu)$, in 
which one observes neutrinos which have not been produced in the ${\pi}$, 
K decays.

\noindent {\bf{Atmospheric neutrinos.}}
Primary cosmic rays produce in the upper atmosphere pions and kaons, which 
decay, $\pi \rightarrow \mu \nu $, $K \rightarrow \mu \nu$; and 
$\mu^\pm \rightarrow \neane_\mu + \neane_e + e^\pm $. Neutrinos are 
produced in a spherical surface, $\sim$ 10 km from the Earth, and 
move towards the Earth.

\begin{figure}
 \vspace{-1.cm}
 \begin{center}
  \mbox{ \epsfysize=7.5cm
         \epsffile{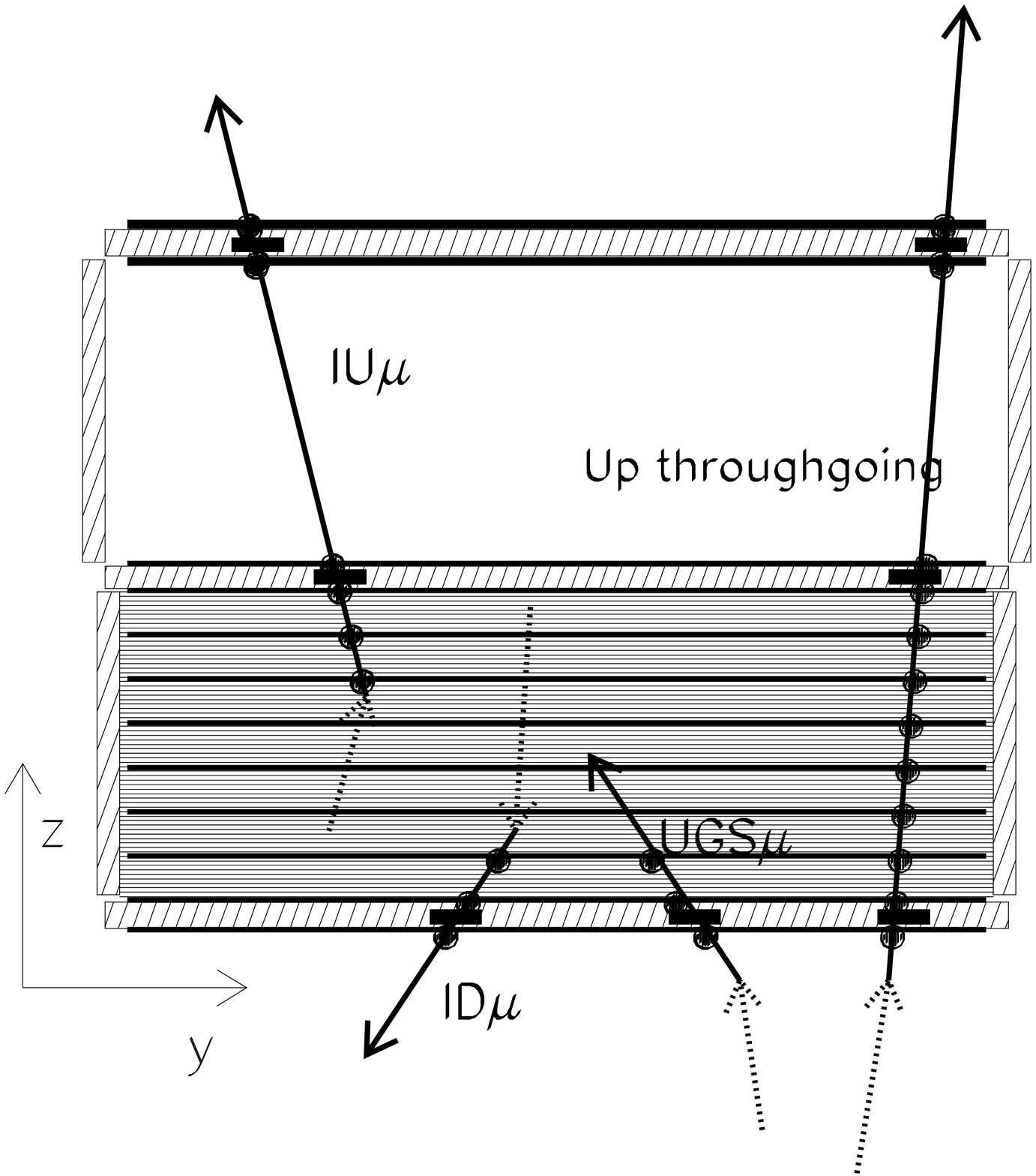} }
 \end{center}
\caption {\label{cross_det}\small Event topologies induced
by $\nu_\mu$ interactions in and around MACRO. The black boxes are scintillator 
hits. 
The track direction and versus are measured by the streamer
tubes and time of flight (ToF).}
\end{figure}

\noindent {\bf{Neutrino-induced Upward Going Muons.}}
They are identified using the streamer tube system (for
tracking) and the scintillator system (for time-of-flight (ToF)
measurement).
A rejection factor of at least $10^5$ is needed in order to separate
the up-going muons from the large background coming from  down-going
muons \cite{atmflu}. Fig. \ref{cross_det} shows  three different topologies of 
neutrino events: up throughgoing muons, semicontained
upgoing muons (IU) and up stopping muons+semicontained downgoing muons (UGS + 
ID) \cite{atmflu}. The parent $\nu_\mu$ energy spectra for the three event
topologies were computed with Monte Carlo methods. The number of events measured
and expected for the three topologies is given in Table \ref{tab:macro}.
Sources of background and systematic effects were studied in detail in
\cite{upgo98} and found to be negligible.

\noindent{\bf High energy data.}
The {\it  up throughgoing muons} come from $\nu_\mu$'s
interacting in the rock below the detector; the $\nu_{\mu}$'s have a median 
energy $\overline E_\nu \sim \ 50\ \GeV$. The muons with $E_\mu > 1\
\GeV$ cross the whole detector; the ToF information
provided by the scintillation counters allows the determination of their
direction (versus).

\begin{figure}
 \vspace{-3.cm}
 \begin{center}
  \mbox{ \epsfysize=7cm
         \epsffile{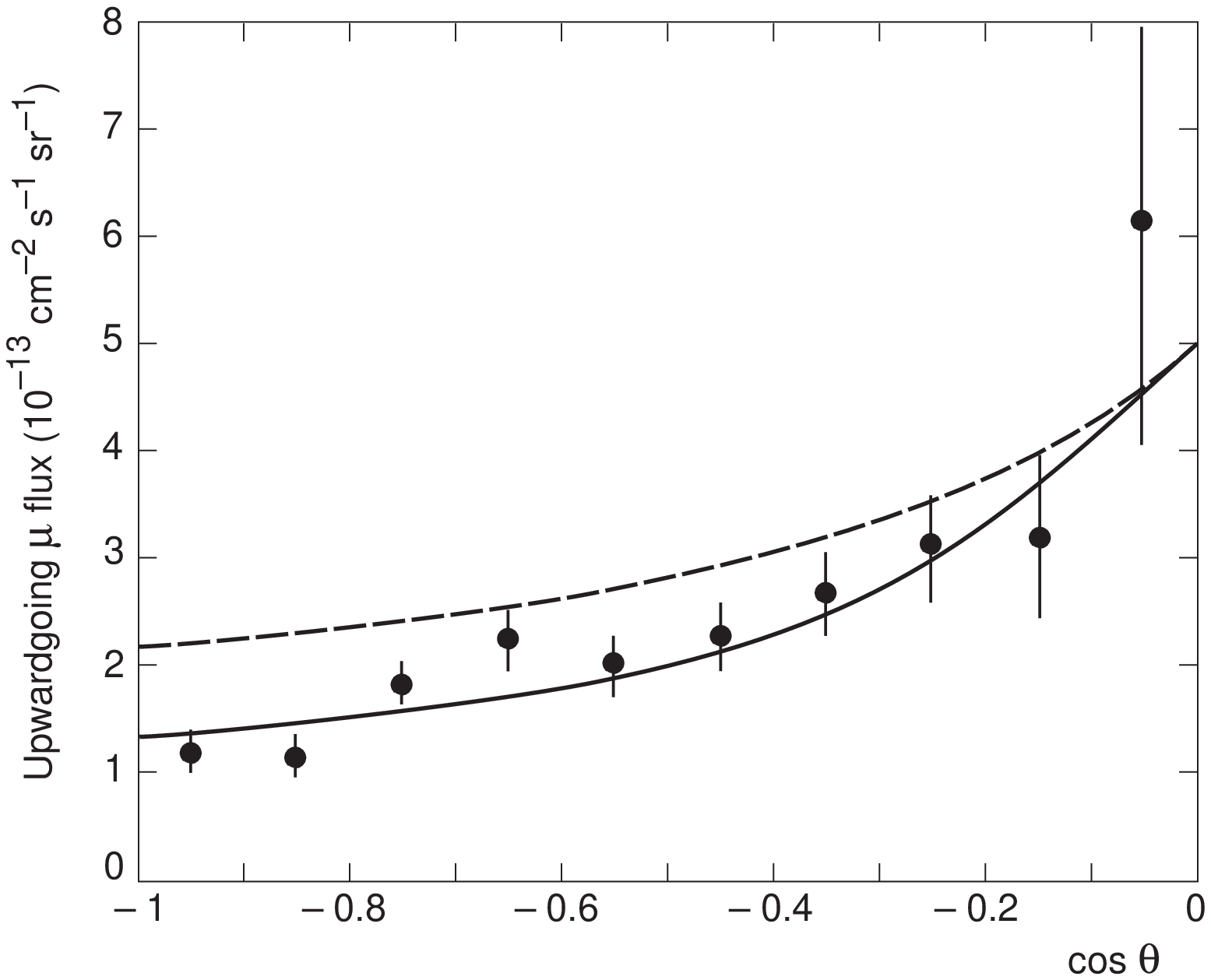}   }
 \end{center}
\vspace{-.3cm}
\caption{\label{flux}\small  The black points are the measured fluxes of the up 
throughgoing muons with $E_{\mu} > 1$ GeV plotted vs. zenith angle $\theta$.
The dashed line is the expectation for no oscillations; it has a 17 \% scale 
uncertainty. The solid 	line is the fit to an oscillated muon flux, obtaining 
maximum mixing and $\Delta m^{2} = 0.0025$ eV$^{2}$.}
\end{figure}

Fig. \ref{flux} 
shows the zenith angle distribution of the measured
flux of up throughgoing muons with energy {\gsim 1 \GeV};
the Monte Carlo expectation for no oscillations is shown as a dashed line,
and for a $\nu_\mu \rightarrow \nu_\tau$ oscillated flux with
$\sin^2 2\theta =1$ and  $\Delta m^2= 0.0025\ \eV ^2$, as a solid line.
The data have been compared with Monte Carlo simulations,
using the neutrino flux computed by the Bartol group and the cross sections 
for  neutrino interactions calculated
using the GRV94 parton distribution; the propagation of muons to the 
detector was  done using the
energy loss calculation by Lohmann {\em et al.} \cite{atmflu}.
The total theoretical uncertainty on the expected muon flux, adding in
quadrature the errors from neutrino flux, cross section and muon propagation,
is 17 \%; it is mainly a scale error that doesn't change the shape of the 
angular distribution.
The  ratio of the observed number of events to the expectation without
oscillations is given in Table \ref{tab:macro}.

\begin{figure}
 \begin{center}
  \mbox{ \epsfysize=7cm
         \epsffile{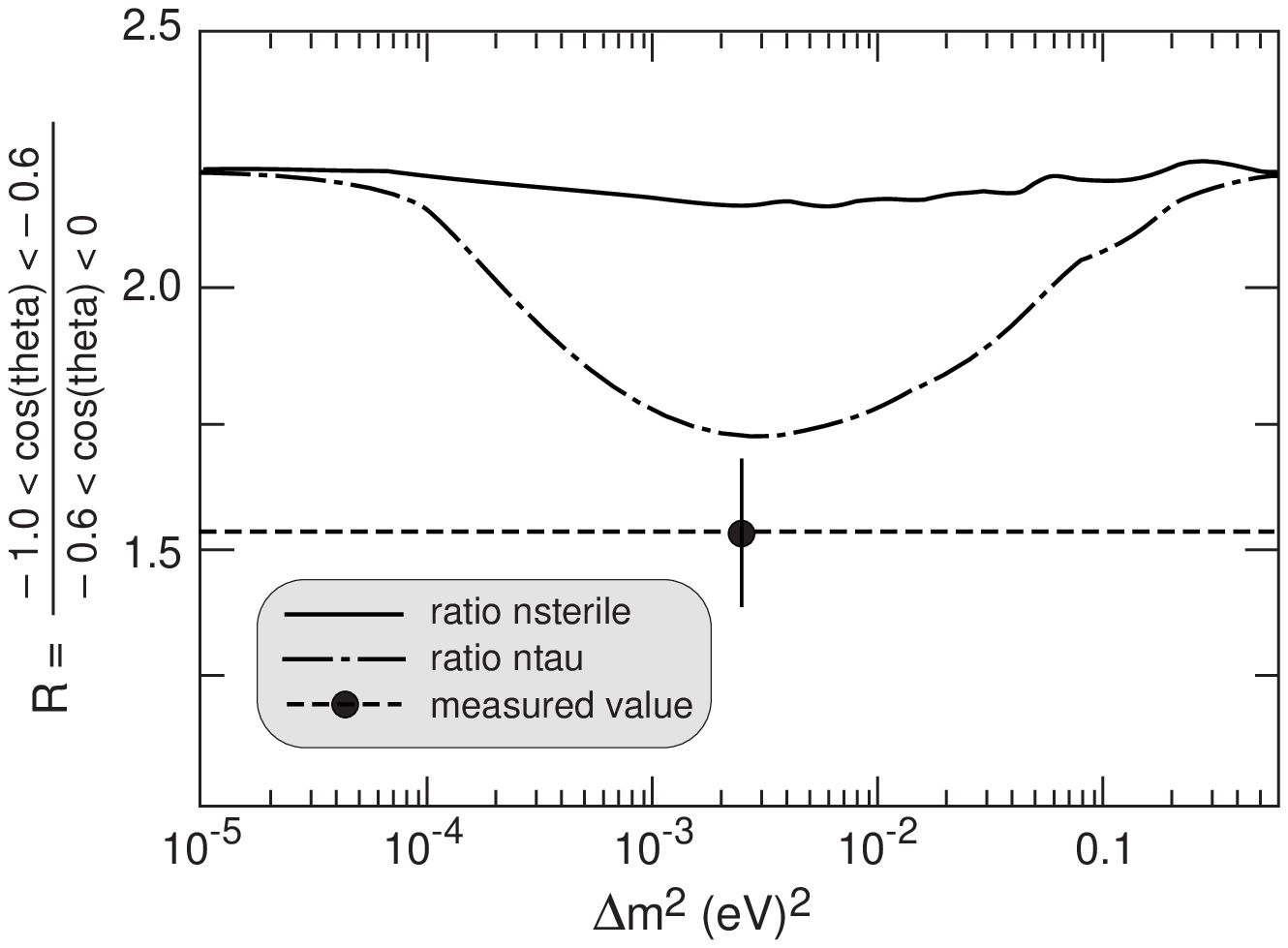} }
 \end{center}
\caption {\label{ratio}\small
The flux ratio vertical/horizontal for upthroughgoing muons.
The measured point is plotted at a $\Delta m^2$ around the minimum of $\chi^2$ for
$\nu_\mu \rightarrow \nu_\tau$ oscillations.}
\end{figure}

The independent probabilities for obtaining the number of events and the 
angular distribution observed have been calculated for various parameter values.
The value of $\Delta  m^2$ obtained from the shape of the
angular distribution is equal to the value
needed to obtain the observed reduction in the number of events; 
for $\nu_\mu \rightarrow \nu_\tau$ oscillations the maximum
probability is 57 \%, with best parameters  $\Delta  m^2 = 0.0025\ \eV ^2$,  
$\sin^2 2\theta = 1$.
The probability for no-oscillations is 0.4 \%.
The probability for $\nu_\mu \rightarrow \nu_{sterile}$ oscillations is 0.15\%;
combining these probabilities with the measured ratio of
horizontal/vertical muons (Fig. \ref{ratio}) 
we conclude that 
$\nu_\mu \rightarrow
\nu_{sterile}$ oscillations are disfavoured compared to $\nu_\mu \rightarrow 
\nu_\tau$.
Fig. \ref{confid}  shows the 90 \% CL regions for $\nu_\mu \rightarrow 
\nu_\tau$, computed according to ref. \cite{feldman}.

\begin{figure}
 \begin{center}
  \mbox{ \epsfysize=8cm
         \epsffile{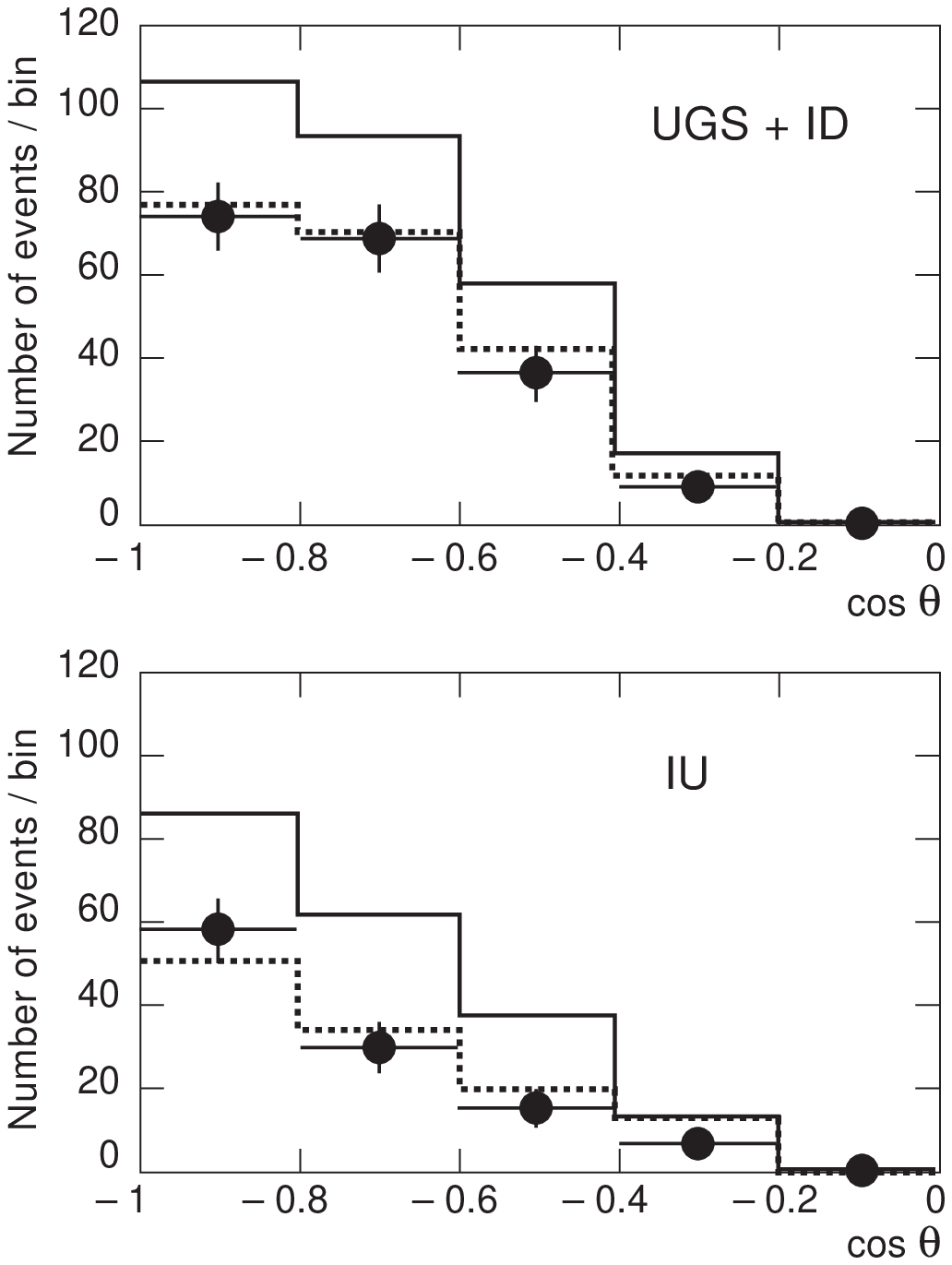} }
 \end{center}
\caption {\label{dist_ang}\small
Measured and expected number of low energy muon neutrino events versus
zenith angle. Top graph: up stopping plus down semicontained;
bottom graph: up semicontained.
The solid lines are the predictions
without oscillations; the dotted lines are the predictions assuming neutrino
oscillations with the parameters obtained from the up throughgoing
sample.}
\end{figure}

\noindent{\bf Low energy data.}
The {\it Upgoing semicontained muons } come from $\nu_\mu$ interactions inside 
the lower apparatus. Since two scintillation counters are intercepted, the 
ToF is applied to identify the upward going muons (Fig. \ref{cross_det}). 
The average parent neutrino energy for these events is  4.2 GeV.
\begin{figure}
 \begin{center}
  \mbox{ \epsfysize=7cm
         \epsffile{confid.eps}   }
 \end{center}
\vspace{-.3cm}
\caption{\label{confid}\small  
Confidence regions for $\nu_\mu \rightarrow \nu_\tau$ oscillations at
the 90 \% CL calculated according to \cite{feldman}, obtained 
from MACRO high energy and low energy data.}
\end{figure}
If the atmospheric neutrino anomalies are the results of
$\nu_\mu \rightarrow \nu_\tau$ oscillations with maximum mixing
and $\Delta m^2$ between $10^{-3}$ and $10^{-2}\ \eV ^2$ one expects a
reduction of about a factor of two in the flux of these events,
without any distortion in the shape of the angular distribution.
This is what is observed in Fig. \ref{dist_ang}.

The {\it up stopping muons}
are due to external $\nu_\mu$ interactions yielding upgoing muon tracks
stopping in the detector; the {\it semicontained downgoing muons} are due to
$\nu_\mu$ induced downgoing tracks with vertices in the lower MACRO
(Fig. \ref{cross_det}).
The events were selected by means of topological criteria; the lack
of time information prevents to distinguish the two sub samples, for which 
an almost equal number of events is expected; the average neutrino energy for 
these events is $\simeq 3.5 \GeV$.
In case of oscillations with the quoted parameters,
a reduction in the flux of the up stopping events
as the semicontained upgoing muons is expected. No reduction is
instead expected for the semicontained downgoing events (coming from
neutrinos with path lengths of $\sim 10 \km$).\par
The MC simulation for the low energy data uses the Bartol neutrino
flux and the neutrino low energy cross sections of ref. \cite{lipari94}.
The number of events and the angular distributions are compared with the
predictions in Fig. \ref{dist_ang} and Table \ref{tab:macro}.
The measured low energy data show a uniform deficit with respect to the 
predictions;
there is good agreement with the predictions based on neutrino
oscillations using the parameters obtained from up throughgoing muons.
The allowed region in the $\Delta  m^2$,  $\sin^2 2\theta $ plane is shown in
Fig. \ref{confid}.\par
\begin{table}
\begin{center}
\begin{tabular}
{cccc}\hline
              & Events   & Predictions (Bartol flux)  &R = Data/MC \\
              & selected & No oscillations            & \\ \hline
Up            & 768      & 989                        & $0.731 \pm 0.028_{st} \pm 0.044_{sys}\pm 0.124_{th}$\\
throughgoing  &          &                            & \\ 
Internal Up   & 135      & 202                        & $0.55 \pm 0.04_{st} \pm 0.06_{sys} \pm 0.14_{th}$\\
              &          &                            & \\ 
Up Stop +     & 229      & 273                        & $0.70 \pm 0.04_{st} \pm 0.07_{sys}
\pm 0.18_{th}$\\
 In Down      &          &                            & \\ \hline
\end{tabular}
\end {center}
\caption {\small Event summary for the MACRO atmospheric neutrino flux 
analyses. The ratios R = Data/MC are relative to MC expectations assuming no 
oscillations (column 3).}
\label{tab:macro}
\end{table}
Using the double ratio
$R= (Data/MC)_{IU} /  (Data/MC)_{ID+UGS}$
between  data and MC of the two low energy data sets, the theoretical
uncertainties on neutrino flux and cross sections
almost disappear (a residual 5 \% uncertainty remains due to the small
differences
between the energy spectra of the two samples).
The average value of the double ratio over the measured zenith angle
distribution is
$R \simeq 0.59 \pm 0.07 stat $.
$R=0.75$ is expected in case of no oscillations, $R=0.58$ in case of 
$\nu_\mu \rightarrow \nu_\tau$  oscillations.

\begin{table}
\begin{center}
\begin{tabular}
{lccccccc}\hline
Source & $\delta$ & Data  & Backg & $\mu$-Flux                & Prev. best                 & $\nu$-Flux  \\
       &          & $3^o$ & $3^o$ & limit                     & $\mu$ limit                & limit       \\
       &          &       &       & $10^{-14}cm^{-2}s^{-1}$   & $10^{-14}cm^{-2}s^{-1}$    & $10^{-6}cm^{-2}s^{-1}$  \\  \hline
SN1987A& $-69.3^o$& 0     & 2.1   & 0.14                      & 1.15 B                     & 0.29                    \\
Vela P & $-45.2^o$& 1     & 1.7   & 0.45                      & 0.78 I                     & 0.84                    \\
SN1006 & $-41.7^o$& 1     & 1.5   & 0.50                      & -                          & 0.92                    \\
Gal. Cen.&$-28.9^o$&0     & 1.0   & 0.30                      & 0.95 B                     & 0.57                    \\
Kep1604& $-21.5^o$& 2     & 1.0   & 1.02                      &         -                  & 1.92                    \\
ScoXR-1& $-15.6^o$& 1     & 1.0   & 0.77                      &      1.5 B                 & 1.45                    \\
Geminga& $18.3^o$ &  0    & 0.5   & 1.04                      & 3.1   I                    & 1.95                    \\
Crab   & $22.o^o$ &  1    & 0.5   & 2.30                      & 2.6 B                      & 4.30                    \\
Her X-1& $35.4^o$ &  0    & 0.2   & 3.05                      & 4.3 I                      & 6.45                    \\
\hline
\end{tabular}
\end {center}
\caption {\small High energy neutrino astronomy: muon and neutrino flux
limits (90 \% CL) for selected sources calculated according to 
the prescription in \cite{feldman}.
Previous best limits (B is for Baksan, I is for IMB) were computed using the 
classical Poissonian method.}
\label{tab:neutrini}
\end{table}

\vspace{0.1cm}
\section{High Energy Muon Neutrino Astronomy}

\begin{figure}
 \begin{center}
  \mbox{ \epsfysize=10cm
         \epsffile{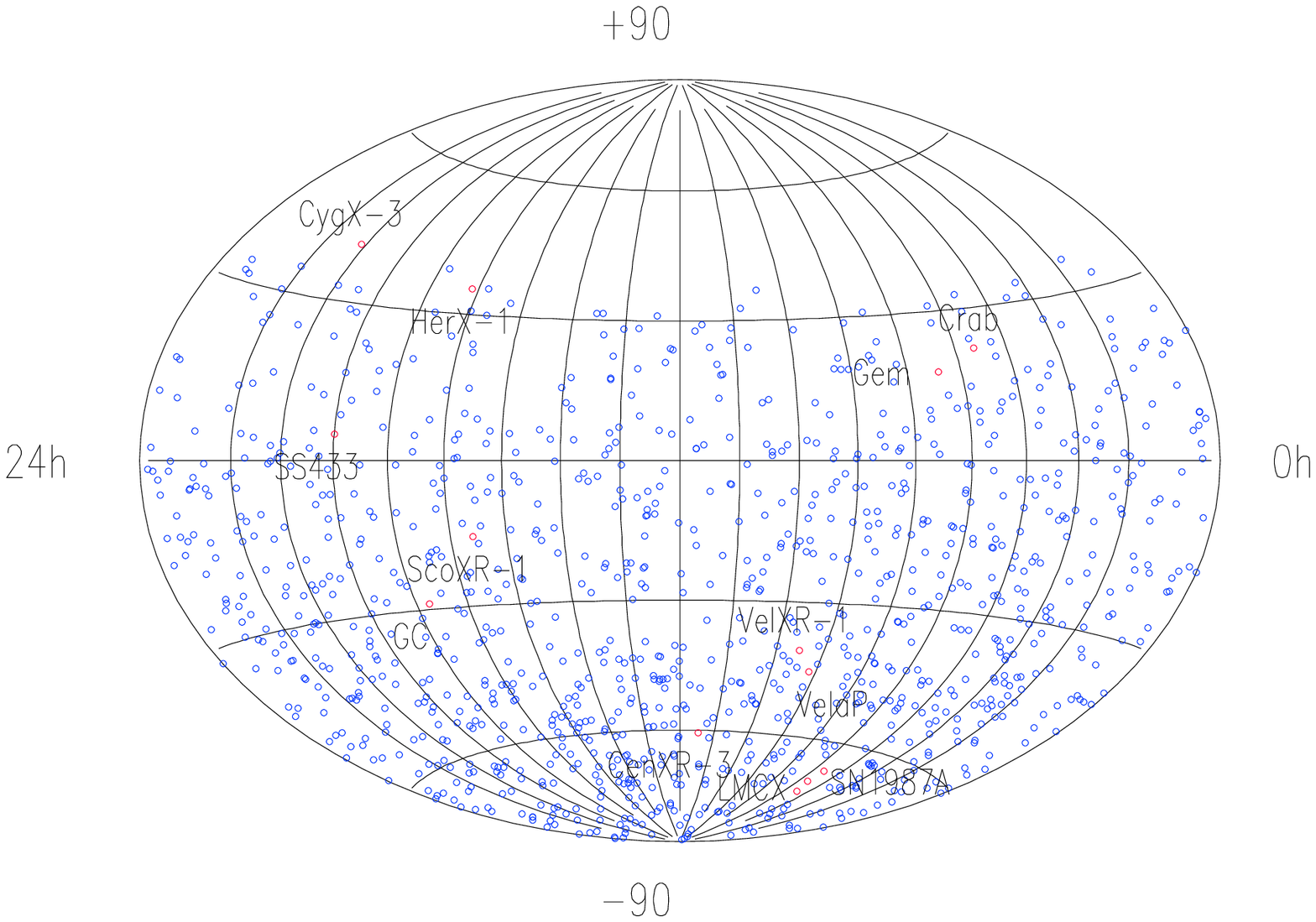} }
\caption{\label{astro}
High energy neutrino astronomy. Upgoing muon event distribution in equatorial
coordinates (1197 events).
}
\end{center}
\end{figure}

The excellent angular resolution of the detector allows a sensitive
search for up-going muons produced by neutrinos coming from celestial
sources; in this case the atmospheric neutrinos are the main background.
The pointing capability of MACRO was demonstrated by the observed "shadows"
of the Moon and of the Sun, which produce a \lq\lq shield\rq\rq~to the cosmic
rays. We used a sample of $ 45 \times 10^6$ downgoing muons, looking at the 
bidimensional density of the events around the center of the Moon and of the 
Sun \cite{mac1}. The density of events shows a depletion with a statistical
significance of $5.5 \ \sigma$ for the Moon and of $4.2 \ \sigma$ for the Sun.
The slight displacements of the maximum deficits are consistent with
the deflection of the primary protons due to the geomagnetic field for the 
Moon and to the combined effect of the magnetic field of the Sun and
of the geomagnetic field. 
The angular resolution, corresponding to a cone angle containing $68 \ \%$
of the events, is $0.9^o \pm 0.3^o$; it is obtained from double muon events.
An excess of events over the atmospheric $\nu$ background was
searched for around the positions of known sources in $3^{\circ}$ (half width)
angular bins. This value was chosen so as to take into account the angular
smearing produced by the multiple scattering in the rock below the detector
and by the energy-integrated angular distribution of the scattered muon,
with
respect to the neutrino direction. A total of 1197 events was used in this 
search, see Fig. \ref{astro}. 
No excess
was observed; the $90 \, \%$ CL limits on
the neutrino fluxes from specific celestial sources are at the level of
$\sim 10^{-6}~{\cm}^{-2}~{\s}^{-1}$, see Table \ref{tab:neutrini} \cite{mac4}.

We also searched for time coincidences of our upgoing muons with $\gamma$-ray
bursts as given in the BATSE 3B and 4B catalogues, for the period April
91 - October 99 \cite{mac4}.
No statistically significant time correlation was found.

\vspace{0.1cm}
\section{Indirect Searches for WIMPs}

\begin{figure}
 \vspace{-2.5cm}
 \begin{center}
  \mbox{ \epsfysize=10cm
         \epsffile{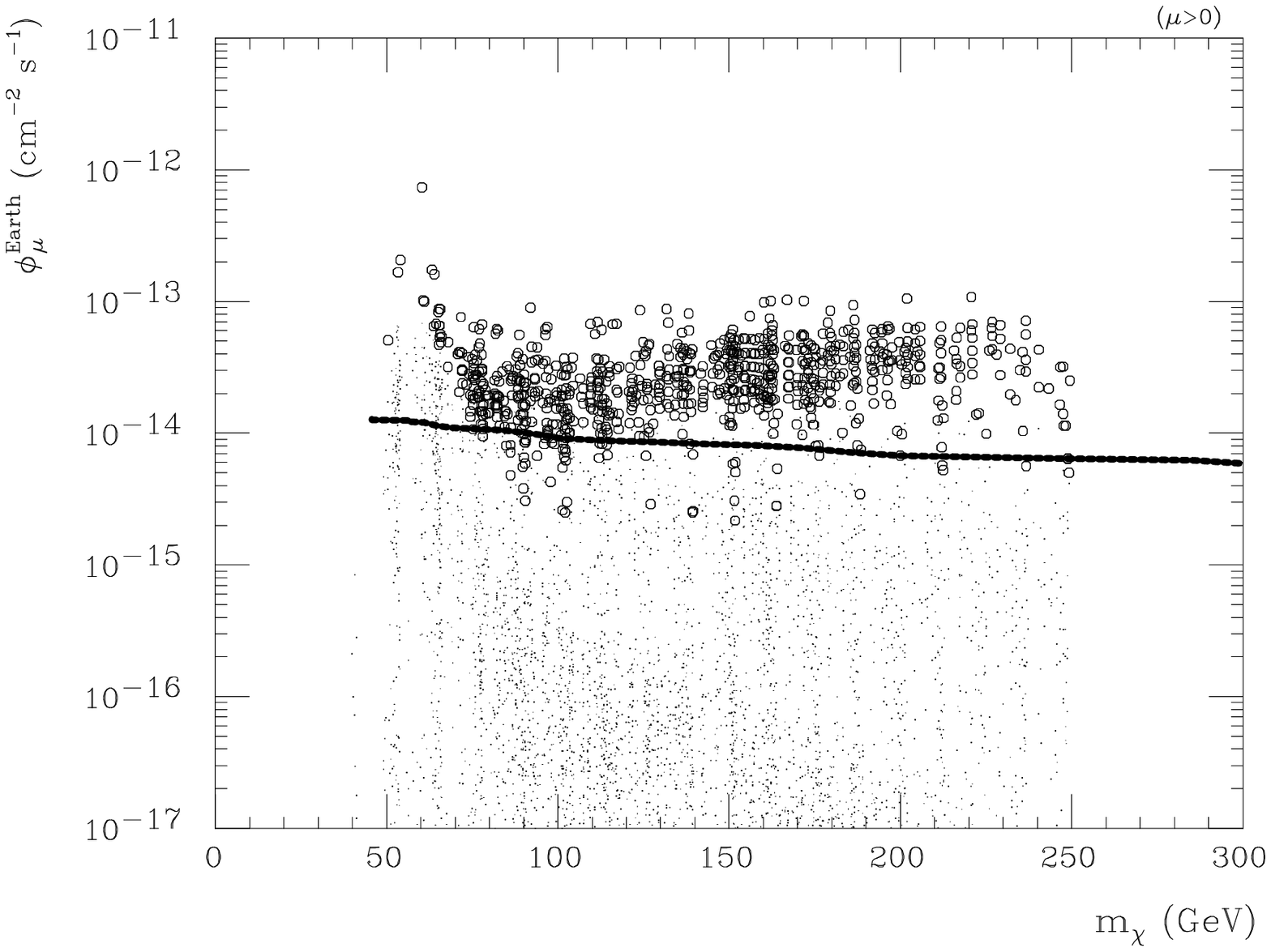}
}
\caption{\label{wimp}
Upward-going muon flux vs neutralino mass $ {m_\chi}$
for ${E_\mu}^{th} = 1 \ \GeV$ from the Earth.
Each dot is obtained varying model parameters. The solid line is the MACRO
flux limit
($90 \, \% \ $CL); the line for the no-oscillation hypothesis is
essentially
indistinguishable in the log scale from the one for
the ${\nu_{\mu} \rightarrow \nu_{\tau}}$ oscillation hypothesis (it could be
about
two times lower). The open circles indicate models {\it excluded} by
direct measurements (particularly
the DAMA/NaI experiment)
and assume a local dark
matter density of ${0.5 \ \GeV \ \cm^{-3}}$.  }
\end{center}
\end{figure}

Weakly Interacting Massive Particles (WIMPs) could be part of the galactic
dark matter; they could be intercepted by celestial bodies, slowed down and
trapped in their centers. WIMPs and anti-WIMPs could annihilate and yield
pions, kaons and then neutrinos of
{\GeV} or
{\TeV} energy coming from a  small angular window from the celestial body  centers.
The $90 \,\%$ CL limit for the flux from the Earth center is $\sim 0.8 \cdot
10^{-14}~{\cm}^{-2}~{\s}^{-1}$
for a $10^{\circ}$ cone around the vertical. The limit from the Sun is 
$\sim 1.4 \times 10^{-14}~{\cm}^{-2}~{\s}^{-1}$.
If the WIMPs are identified with the lowest
mass neutralinos, the MACRO limit constrains the stable
neutralino mass, following the models of Bottino et al., see
Fig. \ref{wimp} 
\cite{mac2}.

\section{Neutrinos from Stellar Gravitational Collapses}

A stellar gravitational collapse (GC) is expected to produce a large burst
of all types of neutrinos and antineutrinos with energies of $ 7-30$~{\MeV} for
a duration of $\sim 10~{\s}$. The ${\bar{\nu}_e}$'s can be detected via 
the reaction
$\bar{\nu}_e~+~p \rightarrow n~+~e^{+}$ in the liquid scintillator.
About $100 \div 150$ ${\bar{\nu}_e}$ events would be detected in our 580 t of
scintillator for a stellar collapse at the center of our Galaxy.
We use two electronic systems for detecting
${\bar{\nu}_e}$'s from stellar gravitational collapses \cite{collapse}.  Both 
systems have an energy
threshold of $\sim 7~{\MeV}$ and record pulse shape, charge and timing
informations. Immediately after a trigger, the dedicated system lowers its 
threshold  to about 1 MeV, for a duration of $800~{\mu\s}$, in order to detect 
(with $\simeq 25 \, \%$ efficiency) the $2.2~{\MeV}$  $\gamma$ released in the 
reaction $n\,+\,p \rightarrow d\,+\,\gamma_{2.2~ \MeV}$ induced by the neutron 
produced in the primary process.
A redundant supernova alarm system is in operation: we have defined a general 
procedure to alert the physics and astrophysics communities in case of an 
interesting alarm \cite{collapse}. A procedure to link the various supernovae observatories
around the world (MACRO, SuperK, LVD, SNO, etc.) was set up. Our
live-time fraction in the last four years was  $\simeq 97.5 \, \%$. No stellar
gravitational collapses were observed
in our Galaxy since $1989$.

\section{Conclusions}

We presented new MACRO data on upgoing muons of low and high energy. These data
are relevant for the study of neutrino oscillations, in particular $\nu_\mu 
\rightarrow \nu_\tau$ \cite{giac-sm99}.

We discussed the limits obtained from the searches for astrophysical sources of 
H.E. neutrinos, for WIMPs and for low energy neutrinos from gravitational stellar 
collapses.

\section{References}
\begin{itemize}
\bibitem[1]{ncim86}C. De Marzo et al., Nuovo Cimento 9C(1986)281; 
M. Calicchio et al., Nucl. Instr. Meth. A264(1988)18;
S. Ahlen et al., Nucl. Instr. Meth. A324(1993)337.\vspace{-0.3cm}
\bibitem[2]{atmflu}M. Ambrosio et al., Phys. Lett. B434(1998)451;
Phys. Lett.  B357(1995)481; Phys. Lett. B478(2000)5.\vspace{-0.3cm}
\bibitem[3]{upgo98}M. Ambrosio et al., Astropart. Phys. 9(1998)105.\vspace{-0.3cm}
\bibitem[4]{feldman} G. Feldman and R. Cousins, Phys. Rev. D57(1998)3873.\vspace{-0.3cm}
\bibitem[5]{lipari94} P. Lipari et al., Phys. Rev. Lett. 74(1995)384.\vspace{-0.3cm}
\bibitem[6]{mac1}M. Ambrosio et al., Phys. Rev. D59(1999)012003.\vspace{-0.3cm}
\bibitem[7]{mac4}M. Ambrosio et al.,``Neutrino astronomy with the MACRO 
detector", astro-ph/0002492.\vspace{-0.3cm}
\bibitem[8]{mac2}M. Ambrosio et al., Phys. Rev. D60(1999)082002.\vspace{-0.3cm}
\bibitem[9]{collapse}M. Ambrosio et al., Astropart. Phys.  8(1998)123.\vspace{-0.3cm}
\bibitem[10]{giac-sm99} G. Giacomelli, Closing lecture at the 1999 S. Miniato 
Workshop, hep-ex/0001008.\vspace{-0.3cm}
\end{itemize}
\end{document}